\documentclass[aps,prl,showpacs,twocolumn,superscriptaddress]{revtex4-2}
\usepackage{bm,color}
\usepackage{graphicx}
\usepackage{amsmath, amssymb}
\usepackage{braket}
\begin{document}
\title{Theory of Collective Excitations in the Quadruple-$Q$ Magnetic Hedgehog Lattices}
\author{Rintaro Eto}
\affiliation{Department of Applied Physics, Waseda University, Okubo, Shinjuku-ku, Tokyo 169-8555, Japan}
\author{Masahito Mochizuki}
\affiliation{Department of Applied Physics, Waseda University, Okubo, Shinjuku-ku, Tokyo 169-8555, Japan}
\date{\today}
\begin{abstract}
Hedgehog and antihedgehog spin textures in magnets behave as emergent monopoles and antimonopoles, which give rise to astonishing transport and electromagnetic phenomena. Using the Kondo-lattice model in three dimensions, we theoretically study collective spin-wave excitation modes of magnetic hedgehog lattices which have recently been discovered in itinerant magnets such as MnSi$_{1-x}$Ge$_{x}$ and SrFeO$_3$. It is revealed that the spin-wave modes, which appear in the sub-terahertz regime, have dominant amplitudes localized at Dirac strings connecting hedgehog-antihedgehog pairs and are characterized by their translational oscillations. It is found that their spectral features sensitively depend on the number and configuration of the Dirac strings and, thus, can be exploited for identifying the topological phase transitions associated with the monopole-antimonopole pair annihilations.
\end{abstract}
\maketitle

\begin{figure}[t]
\centering
\includegraphics[scale=0.5]{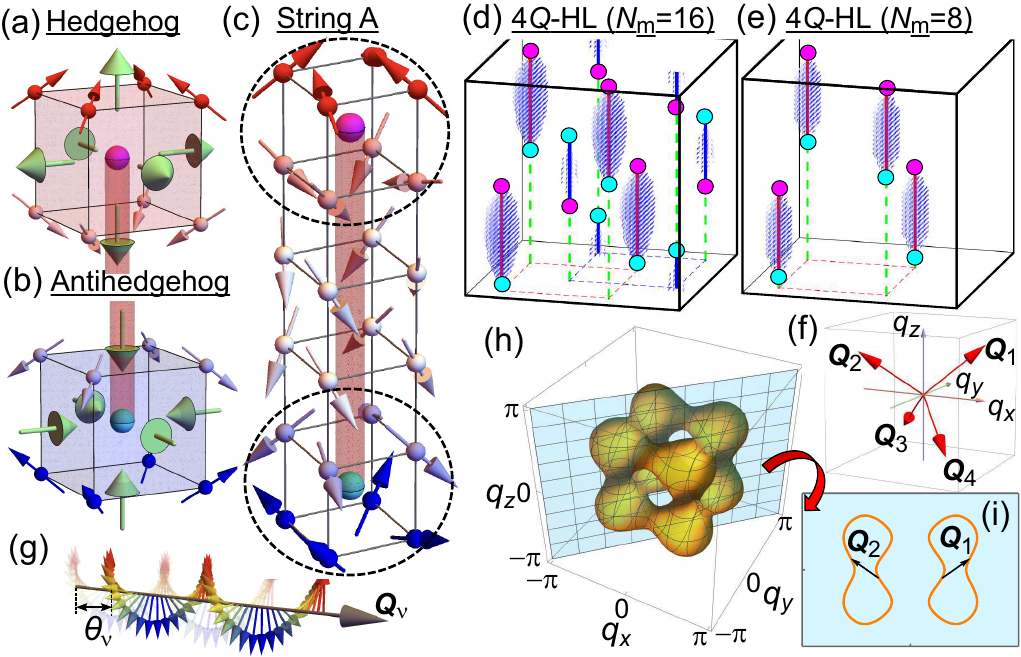}
\caption{(a),~(b) Schematics of hedgehog (a) and antihedgehog (b) spin textures. Green arrows represent emergent magnetic fields around the point defects (Bloch points), and arrows on the cube corners represent the localized spins. (c) Spin configuration of String A. (d),~(e) Spatial distributions of the Bloch points of hedgehogs (magenta) and antihedgehogs (cyan) in the 4$Q$-HLs with (d) $N_{\rm m}$=16 and (e) $N_{\rm m}$=8. Red and blue lines represent the Dirac strings with vorticities of $-1$ and $+1$. (f) Four propagation vectors of magnetic helices constituting the 4$Q$-HLs. (g) Magnetic helix with a propagation vector $\bm Q_\nu$ where $\theta_\nu$ represents the phase degree of freedom with respect to the translation of helix. (h) Fermi surface for the kinetic term of $\mathcal{H}_{\rm KLM}$ in Eq.~(\ref{eq:KLM}). (i) Two-dimensional cross section of the Fermi surface with nesting vectors $\bm Q_1$ and $\bm Q_2$.}
\label{Fig01}
\end{figure}
Magnetic monopoles, elementary particles with isolated magnetic charges in three dimensions, have attracted continuous research interest since Dirac's first proposal in 1931~\cite{Dirac1931}. Although real magnetic monopoles have not been observed in nature, scientists have discovered quasiparticles that virtually behave as magnetic monopoles in condensed-matter systems such as Dirac-electron materials~\cite{Berry1984} and spin-ice pyrochlores~\cite{Castelnovo2012,Pan2016}. Interesting physical phenomena associated with such emergent magnetic monopoles, e.g., fractional excitations~\cite{Fuide2002,Castelnovo2012,Ross2011,Udagawa2019}, anomalous Hall effects~\cite{XuG2011,Uchida2017}, and anomalous Nernst effects~\cite{Liang2016,ZhangH2021}, have been discussed and/or observed in these systems.

Recently, another condensed-matter system that realizes emergent magnetic monopoles was discovered~\cite{Binz2006PRL,Binz2006PRB,ParkJH2011}. It was revealed that an itinerant chiral magnet MnGe hosts a periodic array of magnetic hedgehogs and antihedgehogs called magnetic hedgehog lattice~\cite{Kanazawa2011,Kanazawa2012,Tanigaki2015,Kanazawa2016,Fujishiro2018,Bornemann2019,Kanazawa2020,Fujishiro2021,Pomjakushin2023,Grytsiuk2020,Yaouanc2017,Martin2019,Fujishiro2020,Kanazawa2022,Tokura2021}. These hedgehogs and antihedgehogs can be regarded as magnetic monopoles and antimonopoles, respectively, as they behave as sources and sinks of emergent magnetic fields acting on the conduction electrons via exchange coupling through the Berry-phase mechanism [Figs.~\ref{Fig01}(a)-(c)]. 

The magnetic hedgehog lattice in MnGe is described by a superposition of spin helices with cubic three propagation vectors and, thereby, is referred to as a triple-$Q$ hedgehog lattice (3$Q$-HL). Recently, it was experimentally discovered that substitution of Ge with Si transforms this 3$Q$-HL into another type of hedgehog lattice [Figs.~\ref{Fig01}(d),(e)]. A small-angle neutron-scattering experiment revealed that the hedgehog lattice in MnSi$_{1-x}$Ge$_{x}$ is characterized by a superposition of spin helices with tetrahedral four propagation vectors [Fig.~\ref{Fig01}(f)], i.e., the quadruple-$Q$ hedgehog lattice (4$Q$-HL)~\cite{Fujishiro2019}.  The 4$Q$-HL was observed also in the perovskite ferrite SrFeO$_3$~\cite{Ishiwata2020}. 

In fact, there are several kinds of 4$Q$-HL states with different number of hedgehogs and antihedgehogs in the magnetic unit cell because the spin structure of 4$Q$-HLs are characterized not only by the propagation vectors $\bm Q_\nu$ ($\nu$=1-4)~\cite{Okumura2020,Kitaori2021,Shimizu2021} but also by other internal degrees of freedom, e.g., net magnetization~\cite{Okumura2020}, chirality~\cite{Okumura2022}, and relative phases $\theta_\nu$ of the superposed spin helices [Fig.~\ref{Fig01}(g)]~\cite{Shimizu2022,Hayami2021NComm}.
Therefore, we expect that MnSi$_{1-x}$Ge$_{x}$ and SrFeO$_3$ offer precious opportunities to study the tunability of transport and thermoelectric properties via the field-induced variation of topological nature of the hedgehog lattices~\cite{Shimizu2022,Hayami2021JPCM,Kato2023}. 

The dynamical phenomena associated with these emergent magnetic monopoles are more interesting. From the electromagnetic duality in Maxwell's equations, the dynamics of magnetic monopoles is expected to induce an electric field, as the dynamics of electric charge induces a magnetic field. Through inducing the electric field while moving, the emergent monopoles behaves as dyons, i.e., hypothetical particles in the grand unified theory~\cite{Schwinger1969}. Novel transport phenomena and optical/microwave device functions are expected to emerge from such emergent dyons~\cite{ZhangX2016,ZouJ2020,Kato2021,Paradezhenko2022,Aoki2023}. 
In order to study new physical phenomena and device functions originating from the dynamics of such emergent monopoles and dyons, knowledge of their intrinsic excitations is essentially important.

In this Letter, we theoretically study collective excitation modes of the 4$Q$-HLs in itinerant chiral magnets using a microscopic spin-charge coupled model in three dimensions. We construct a model on the chiral-cubic lattice so as to reproduce the distinct two types of 4$Q$-HLs with very short periods of a few atomic sites, which are realized in MnSi$_{1-x}$Ge$_{x}$ and SrFeO$_3$, based on the microscopic insights into these materials.
We first reveal that the nesting of Fermi surfaces can work as a principal mechanism for stabilizing the 4$Q$-HL textures. Then we find that the 4$Q$-HLs have characteristic collective excitation modes associated with translational motion of Dirac strings. Because these collective excitation modes are dominated by the Dirac strings, presence or absence of the modes sensitively depends on the number of monopole-antimonopole pairs and their spatial configuration. Indeed, we demonstrate that a certain mode disappears upon a topological phase transition associated with vanishing of Dirac strings due to the field-induced monopole-antimonopole pair annihilations. This finding will contribute to characterization of the magnetic topologies in hedgehog-hosting magnets and will be a basis for studying the monopole-induced physical phenomena in matters.

We start with the Kondo-lattice model on a cubic lattice with the Dzyaloshinskii-Moriya (DM) interactions, which describes the coupling between the conduction electrons and the localized spins in itinerant chiral magnets, and some additional terms. The Hamiltonian is given by 
$\mathcal{H}=\mathcal{H}_{\rm KLM}+\mathcal{H}_{\rm Zeeman}+\mathcal{H}_{\rm local}$
with
\begin{align}
&\mathcal{H}_{\rm KLM}=
-\sum_{i,j,\sigma} t_{ij}\hat{c}_{i\sigma}^\dagger\hat{c}_{j\sigma}
-J_{\rm K} \sum_{i}\hat{\bm s}_i\cdot{\bm S}_i, 
\label{eq:KLM} \\
&\mathcal{H}_{\rm Zeeman}=
-\left[\bm H + \bm h(t) \right] \cdot \sum_i \bm S_i, 
\label{eq:Zeeman} \\
&\mathcal{H}_{\rm local}=
 \sum_{<i,j>}J_{\rm AFM}\bm S_i \cdot \bm S_j 
-D\sum_{<i,j>}\bm e_{ij} \cdot (\bm S_i \times \bm S_j),
\label{eq:Local}
\end{align}
where $\hat{c}_{i\sigma}^\dagger$ ($\hat{c}_{i\sigma}$) denotes a creation (annihilation) operator of a conduction electron with spin $\sigma(=\uparrow, \downarrow)$ at site $i$. The first term of $\mathcal{H}_{\rm KLM}$ describes the kinetic energy of the conduction electrons with the nearest-neighbor hopping $t_1(=1)$ and the fourth-neighbor hopping $t_4(=-1)$. The second term of $\mathcal{H}_{\rm KLM}$ describes the Kondo exchange coupling between the localized classical spins $\bm S_i$ $(|\bm S_i|=1)$ and the conduction-electron spins $\hat{\bm s}_i=\left(1/2\right) \sum_{\sigma\sigma'}\hat{c}_{i\sigma}^\dagger{\bm \sigma}_{\sigma\sigma'}\hat{c}_{i\sigma'}$ where $\bm \sigma$ represents the vector of Pauli matrices. The term $\mathcal{H}_{\rm Zeeman}$ denotes Zeeman coupling associated with the external magnetic field, while the term $\mathcal{H}_{\rm Local}$ describes the antiferromagnetic (AFM) exchange interactions and the DM interactions between the nearest-neighbor localized spins, where ${\bm e}_{ij}$ represents the bond-directional unit vectors. These interactions originate from the hybridizations among the localized orbitals and the spin-orbit coupling~\cite{Nomoto2020}. We set $J_{\rm K}$=0.8 and $J_{\rm AFM}$=0.0008, and we examine both the cases with and without DM interactions, i.e., $D=0$ and $D=0.0002$.

The 4$Q$-HL texture is described by a superposition of four magnetic helices governed by the nested Fermi surfaces of $\mathcal{H}_{\rm KLM}$ [Figs.~\ref{Fig01}(h) and (i)]. More specifically, the propagation vectors of helices $\{\bm Q_\nu\}$ coincide the nesting vectors [Fig.~\ref{Fig01}(i)]. This is because the effective two-body exchange interactions between the localized spins are governed by the bare susceptibility $\chi^0_{\bm q}$, which has maxima at momenta $\bm q$ that correspond to the nesting wavenumbers~\cite{Hayami2017}. Indeed, when $J_{\rm K}$ is weak, the second-order perturbation theory gives $-(J_{\rm K}/2)^2 \sum_{{\bm q}\in{\rm BZ}} \chi^0_{\bm q}|\bm S_{\bm q}|^2$ as the effective interaction.

The tetrahedral four propagations vectors $\{\bm Q_\nu\}$ are represented by $\bm Q_1=(Q_{\rm abs},Q_{\rm abs},Q_{\rm abs})$, $\bm Q_2=(-Q_{\rm abs},-Q_{\rm abs},Q_{\rm abs})$, $\bm Q_3=(-Q_{\rm abs},Q_{\rm abs},-Q_{\rm abs})$, and $\bm Q_4=(Q_{\rm abs},-Q_{\rm abs},-Q_{\rm abs})$. In the present study, we set the chemical potential $\mu=-3.79$, which gives $Q_{\rm abs} \approx \pi/4$. This wavenumber corresponds to a spatial period of $\lambda=2\pi a/\sqrt{3}Q_{\rm abs}\sim$2.15 nm, if we assume the lattice constant of $a=0.465$ nm, which reproduces the experimentally observed $\lambda=$1.9-2.1 nm for MnSi$_{1-x}$Ge$_{x}$~\cite{Fujishiro2019,Fujishiro2020}. Note that the following results are robust, and a fine tuning of the chemical potential is not required~\cite{SM}. 

\begin{figure}[t]
\centering
\includegraphics[scale=0.5]{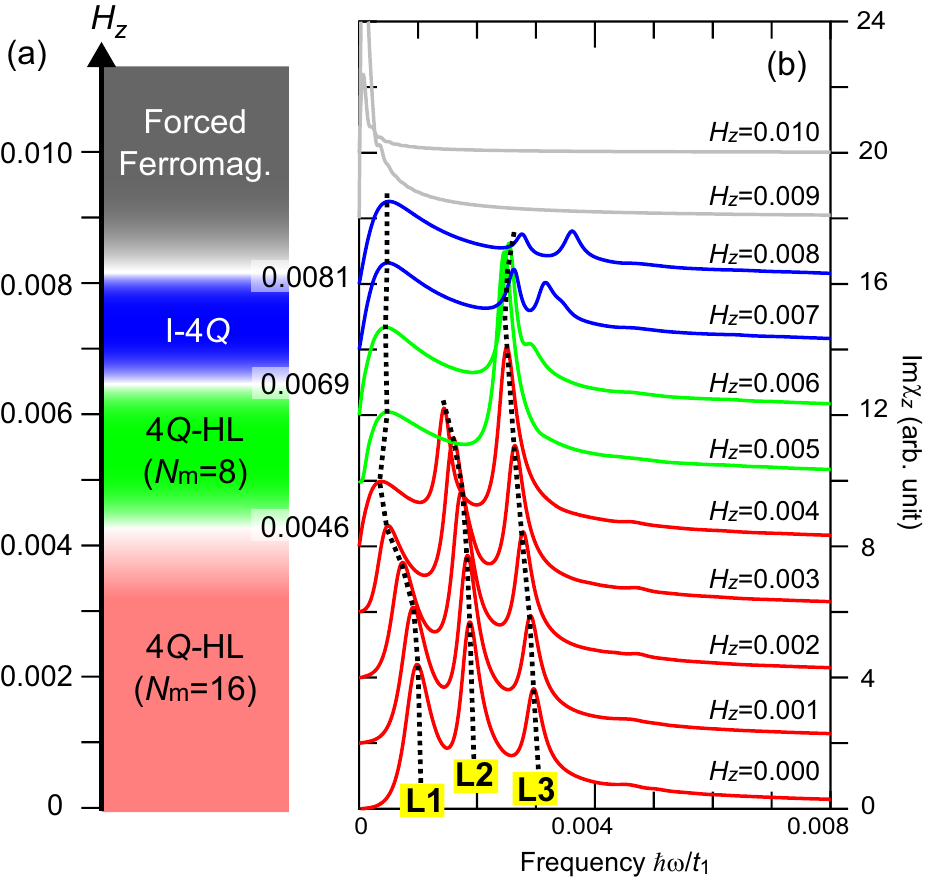}
\caption{(a) Ground-state phase diagram as a function of the external magnetic field $H_z$, which contains the 4$Q$-HL with $N_{\rm m}$=16, another 4$Q$-HL with $N_{\rm m}$=8, the intermediate 4$Q$ (I-4$Q$) phase, and the forced ferromagnetic phase. (b) Imaginary part of the longitudinal dynamical magnetic susceptibility for various values of $H_z$. The DM parameter is set to be $D=0.0002$.}
\label{Fig02}
\end{figure}
We investigate the ground-state phase diagram of the Hamiltonian $\mathcal{H}$ by using the adiabatic Landau-Lifshitz-Gilbert equation, $\dot{\bm S}_i = -{\bm S}_i\times{\bm H}_i^{\rm eff} + \alpha_{\rm G}{\bm S}_i\times\dot{\bm S}_i$, where $\alpha_{\rm G}$ is the Gilbert-damping coefficient. The effective local magnetic field is calculated by the first-order differential of thermodynamic potential as $\bm H_i^{\rm eff}=-\partial\Omega/\partial \bm S_i$. For this formulation, we assume an adiabatic condition that dynamics of the conduction electrons are sufficiently fast compared to that of the localized spins and follow them smoothly, so that the conduction electrons are always in equilibrium. The calculations are numerically performed by using the kernel polynomial method combined with the automatic differentiation technique based on the chain rule of Chebyshev polynomials \cite{Wang1994,Motome1999,Weisse2006,Barros2013,Chern2018}. This method is known to be powerful when simulating the low-energy dynamics in spin-charge coupled systems~\cite{Wang2016,Chern2018,Eto2022}. 

The phase diagram obtained for a system of $16^3$ sites is shown in Fig.~\ref{Fig02}(a). We also confirm that the phase diagram is not changed in an analysis of a larger size system with $32^3$ sites. Here the DM parameter is set to be $D=0.0002$. When $H_z$=0, a 4$Q$-HL state with $N_{\rm m}=16$ appears where $N_{\rm m}$ is the total number of hedgehogs and antihedgehogs in the magnetic unit cell. It is worth mentioning that the stable 4$Q$-HL state depends on the presence or absence of the DM interaction~\cite{Okumura2020,Okumura2022}. When the DM interaction is absent ($D$=0), the 4$Q$-HL consisting of two right-handed helices and two left-handed helices is stabilized. On the contrary, the 4$Q$-HL consisting of four-righthanded helices is stabilized in the presence of the DM interaction ($D=0.0002$). In the following, we focus on the 4$Q$-HL in the latter case. This chiral 4$Q$-HL state contains two inequivalent Dirac strings (A and B), each of which connects a hedgehog and an antihedgehog along the $z$ axis. There are eight strings in total, four each for A and B. Note that the magnitude of DM interaction required for stabilizing the chiral 4$Q$-HL is as small as $D=0.0002$, which is approximately 35\% of $J_1^{\rm eff}$ and 25\% of $J_{\rm AFM}$. Here $J_1^{\rm eff}=-(J_{\rm K}/2)^2(1/N_{\bm q})\sum_{{\bm q}\in{\rm BZ}}\chi^0_{\bm q}e^{i{\bm q}\cdot{\bm e}_\gamma}$ $(\gamma=x,y,z)$ is the coupling constant of the effective nearest-neighbor ferromagnetic exchange interactions mediated by the conduction electrons.

When the external magnetic field is absent ($H_z=0$), the Dirac strings A and B have the same length $d_{\rm A}=d_{\rm B}=4a$. As $H_z$ is increased, both $d_{\rm A}$ and $d_{\rm B}$ decrease, keeping the relationship $d_{\rm A}\geq d_{\rm B}$. The length is defined by $d=|\bm R_{\rm H}-\bm R_{\rm AH}|$. Here $\bm R_{\rm H}$ ($\bm R_{\rm AH}$) is a center position of the unit cube on which the (anti)hedgehog is defined, which takes integer values with the lattice constant $a$ as the unit. When the field strength reaches $H_z\approx$0.0046, the length $d_{\rm B}$ reaches zero first, which results in the pair annihilations of the hedgehog and antihedgehog connected by the String B and vanishing of the String B. Consequently, a topological phase transition occurs from the 4$Q$-HL state with $N_{\rm m}$=16 to that with $N_{\rm m}$=8. This phase transition is characterized by a change of the phase $\Theta$, which is defined by $\Theta=\pi-\left|\pi-{\rm mod}\left[\Theta^\prime,2\pi\right]\right|$ with $\Theta^\prime=\sum_{\nu=1}^4 \theta_\nu$, where $\theta_\nu$ is the relative phase shift of helix with the propagation vector $\bm Q_\nu$ [Fig.~\ref{Fig01}(g)]. With increasing $H_z$, $\Theta$ decreases from $\Theta=\pi/3$ and is strongly suppressed around the phase boundary. While $\Theta$ is still finite at the phase boundary, it becomes zero when the system goes a little inside the $N_{\rm m}$=8 phase [Fig.~\ref{Fig01}(e)] \cite{Shimizu2022}. The ellipticity $\varepsilon_\nu$ also exhibits a characteristic change with a minimum around the phase boundary~\cite{SM}.

\begin{table}[tb]
\centering
\caption{Properties of the collective excitation modes in the 4$Q$-HL states. The second column shows the Dirac strings at which the mode has large amplitude, while the third column shows the relevant 4$Q$-HL state.}
\label{table:eigenmodes}
\centering
\begin{tabular}{cccc}
\hline
Mode \quad & \quad Relevant Dirac strings \quad & \quad Relevant 4$Q$-HL state(s) \\
\hline
\hline
L1 & -         & $N_{\rm m}=16$, $N_{\rm m}=8$ \\
L2 & Strings B & $N_{\rm m}=16$                \\
L3 & Strings A & $N_{\rm m}=16$, $N_{\rm m}=8$ \\
\hline
\end{tabular}
\end{table}
Now we study the collective excitation modes in these chiral 4$Q$-HL phases by calculating the longitudinal dynamical magnetic susceptibility $\chi_z(\omega)$. Time profiles of the total magnetization $S_z(t)$ are calculated using the aLLG equation after applying a short-time pulse of $\bm h(t)=h_z\delta(t)\bm e_z$ and perform the Fourier transformation of $\Delta S_z(t)=S_z(t)-S_z(0)$ to obtain $\chi_z(\omega)$. The Gilbert-damping coefficient is fixed at $\alpha_{\rm G}$=0.04 for the simulations. The obtained spectra of Im$\chi_z(\omega)$ show that three intrinsic excitation modes (L1, L2, and L3) appear in the $N_{\rm m}=16$ phase, whereas the mode L2 disappears when the system enters the $N_{\rm m}=8$ phase. Note that these modes appear in the sub-terahertz regime because $\omega$=0.004 in Fig.~\ref{Fig02}(b) corresponds to 1 THz approximately when we assume $t_1=1$ eV.

\begin{figure}[tb]
\centering
\includegraphics[scale=1.0]{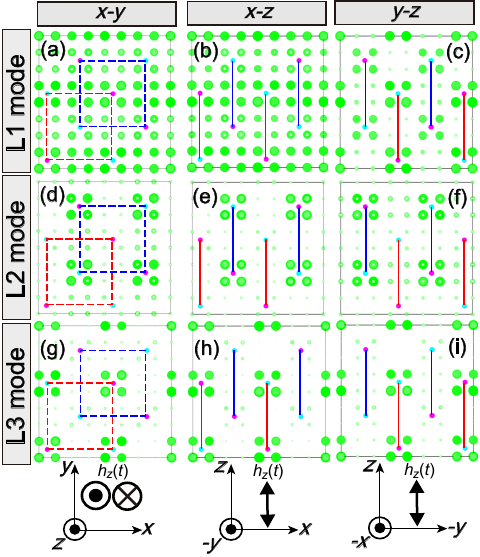}
\caption{(a)-(c) Spatial maps of the oscillation amplitudes projected onto the $xy$, $yz$ and $zx$ planes for the L1 mode in the 4$Q$-HL phase with $N_{\rm m}=16$ when $H_z=0.001$ and $D=0.0002$. (d)-(f) Those for the L2 mode. (g)-(i) Those for the L3 modes. The red (blue) solid lines represent Strings A (B), while the red (blue) dashed lines connect Strings A (B) to show their spatial configuration seen along the $c$ axis, which are also shown in Figs.~\ref{Fig01}(d) and \ref{Fig01}(e).}
\label{Fig03}
\end{figure}
A notable property of these longitudinal modes is their spatial localization. Figures~\ref{Fig03}(a)-(i) show the spatial distribution of oscillation amplitudes for the L1, L2, and L3 modes in the $N_{\rm m}=16$ phase at $H_z=0.001$ excited by a time-dependent field $\bm h(t)=h_z\sin(\omega_{\rm ac}t)\bm e_z$ with a small amplitude of $h_z=10^{-5}$. Here the frequency $\omega_{\rm ac}$ is fixed at the eigenfrequency of the corresponding mode, and we used $\alpha_{\rm G}=0.008$ for the simulations. Similar plots as Fig.~\ref{Fig03} for the modes in the $N_{\rm m}=16$ phase at $H_z=0$ and those for the modes in the $N_{\rm m}=8$ phase at $H_z=0.005$ are presented in Figs. S4 and S5 in the Supplemental Materials~\cite{SM}, respectively. Here sizes of the green balls represent norms of the oscillation amplitudes $\left|\delta{\bm S}_i\right|=\left[ \sum_{\mu=x,y,z} \left(\delta S_{i\mu}\right)^2 \right]^{1/2}$ during a period of the oscillation. The ball sizes are normalized with respect to the largest value in the unit cell after subtracting a spatially uniform component for clear visibility.

In these figures, we find that the L2 mode is localized at the Strings B, whereas the L3 mode is at the Strings A. Consequently, the L2 mode disappears when the system enters the $N_{\rm m}=8$ phase because the Strings B disappear as the hedgehog-antihedgehog pair annihilations occur upon this phase transition [Fig.~\ref{Fig02}(b)]. The disappearance of L2 mode manifests the change in magnetic topology from $N_{\rm m}=16$ to $N_{\rm m}=8$. This conversely indicates that this topological transition can be observed in the magnetic resonance spectra. On the contrary, the L3 mode survives after this phase transition [Fig.~\ref{Fig02}(b)].  
With further increasing $H_z$, the system enters the intermediate 4$Q$ (I-4$Q$) phase and subsequently the forced ferromagnetic (FFM) phase. In the I-4$Q$ phase, the hedgehog and antihedgehog connected by a Dirac string collide to be merged with their cores sharing the same cubic unit cell~\cite{SM}. In this phase, there still exists nonzero scalar spin chirality as a residue of Strings A, although it is no longer quantized, and the L3 mode survive because of the remnant of topology. The mode disappears upon the transition to the collinear FFM phase at $H_z \approx 0.0081$, in which the scalar spin chirality vanishes~\cite{SM}.

In contrast to the L2 and L3 modes, the L1 mode is not localized, but its oscillation amplitude is widely distributed in the magnetic unit cell. We also mention two aspects of the spin-wave modes. First, the L1 mode is not $C_{4z}$-symmetric with respect to the spatial distribution of the oscillation amplitudes as seen in Figs.~\ref{Fig03}(b) and \ref{Fig03}(c) because of the  reduced symmetry of the spin texture~\cite{SM}. Second, the three spin-wave modes are gapped when $H_z$=0 and distinct from those in the 3$Q$ hedgehog lattices discussed in  Ref.~\cite{ZhangX2016}, which are gapless at zero field.

\begin{figure}[tb]
\centering
\includegraphics[scale=0.5]{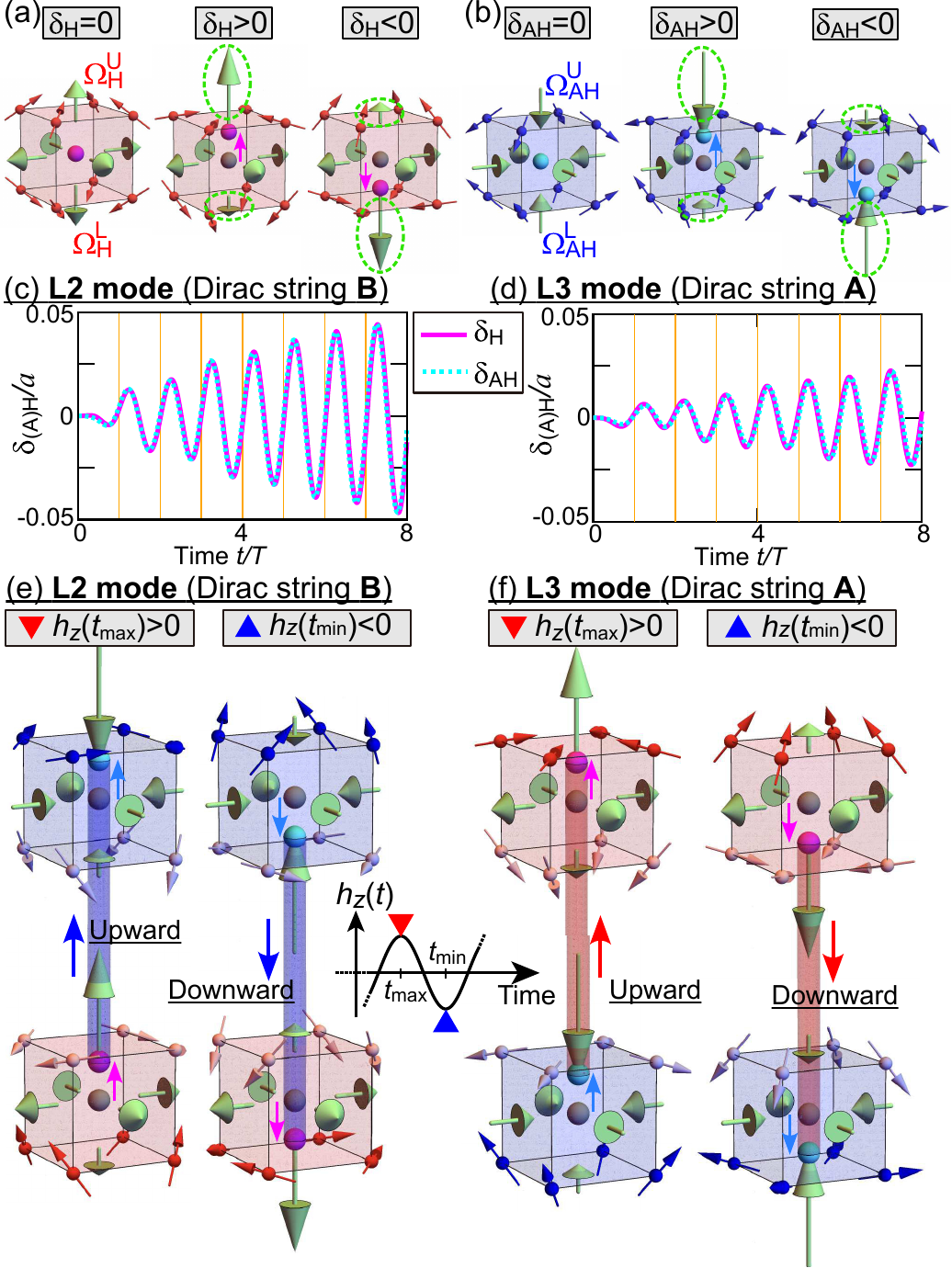}
\caption{(a),(b) Relationship of the positions of Bloch points in the crystallographic unit cell and local solid angles composed of localized spins. The length of the green arrows represent the magnitude of solid angle spanned by four localized spins, which is roughly proportional to the local emergent magnetic fields. (c) Time profiles of the displacements $\delta_{\rm (A)H}$ of the Bloch points for the String B in the L2 mode.  (d) Time profiles of $\delta_{\rm (A)H}$ for the String A in the L3 mode. (e),(f) Schematics of the oscillatory translational motion of the Dirac strings. $T=2\pi/\omega_{\rm ac}$ is the time periodicity of the ac magnetic field.}
\label{Fig04}
\end{figure}
To clarify the characteristics of these intrinsic collective modes, we first focus on a cubic cell with eight spins at its corners that constitute a hedgehog or an antihedgehog spin texture. The center of the (anti)hedgehog is located within this cube. In the absence of external magnetic field, the (anti)hedgehog spin texture is $C_{2x}$-, $C_{2y}$-, and $C_{2z}$-symmetric, and thus the center of mass of the solid angles spanned by four-spin plaquettes on the six cube faces coincides with the cube center. On the contrary, under application of a magnetic field along $z$-axis, the spin texture is no longer $C_{2x}$- and $C_{2y}$-symmetric, and, thereby, the center of mass of the spin solid angles deviates from the cube center along the $z$ axis. 

When a magnetic field along the $z$ axis is applied to the (anti)hedgehog spin texture on a cube, the solid angles $\Omega_{\rm (A)H}^{\rm U}$ and $\Omega_{\rm (A)H}^{\rm L}$ spanned, respectively, by the four-spin plaquette on the upper face and that on the lower face of the cube become different, whereas the four-spin plaquettes on the side faces constitute solid angles of the same magnitude according to the symmetry. Consequently, the center of mass of the spin solid angles shifts along the $z$ axis as shown in Figs.~\ref{Fig04}(a) and (b). We define the displacement $\delta$ as 
$\delta_{\rm (A)H}/a \equiv \frac{1}{2}(|\Omega_{\rm (A)H}^{\rm U}|-|\Omega_{\rm (A)H}^{\rm L}|)/(|\Omega_{\rm (A)H}^{\rm U}|+|\Omega_{\rm (A)H}^{\rm L}|)$ with $a$ being the lattice constant. 

When the magnetic field temporally oscillates along the $z$ axis, the center of mass of the spin solid angles dynamically changes its position along the $z$ axis, i.e., moves upwards ($\delta_{\rm (A)H}>0$) and downwards ($\delta_{\rm (A)H}<0$) in an oscillatory manner. The temporal displacement of the center of mass can be calculated by simulating the time evolution of localized spins when the mode is excited by ac magnetic field $\bm h(t)$ with the corresponding frequency. In Figs.~\ref{Fig04}(c) and \ref{Fig04}(d), the calculated time profiles of displacements $\delta_{\rm (A)H}$ are plotted for the String B in the L2 mode and the String A in the L3 mode, respectively. We find that the hedgehogs and antihedgehogs oscillate with the same phase. Such an in-phase oscillation of the center-of-mass of each (anti)hedgehog can be regarded as coherent translational motion of the Dirac strings. Schematics of the corresponding motion are also shown in Figs.~\ref{Fig04}(e) and \ref{Fig04}(f).

The oscillatory variation of solid angles of four-spin plaquettes orthogonal to the $z$ axis is analogous to the breathing mode in a skyrmion crystal~\cite{Mochizuki2012}. It is known that a skyrmion crystal has one breathing mode because it is crystallographically composed of one skyrmion. In contrast, the present 4$Q$-HL states have multiple modes because it contains crystallographically inequivalent two Dirac strings with different vorticities of $\pm1$. In this case, two localized modes, i.e., the String-A activated (L3) and the String-B activated (L2) modes are possible. The number of modes active to the longitudinal field is governed by the number of crystallographically inequivalent Dirac strings in the hedgehog-antihedgehog lattice states. This indicates that the absorption spectra in the sub-terahertz regime provide a strong clue to identifying real-space configurations of the hedgehog lattice states and their topological phase transitions caused by the monopole-antimonopole pair annihilations.

In summary, we have theoretically studied the spin-wave modes of the 4$Q$-HLs in the spin-charge coupled metallic magnets described by the chiral Kondo-lattice model. We have discovered translation modes of Dirac strings in the hedgehog lattices. It has been found that number of the collective modes is governed by the number of crystallographically inequivalent Dirac strings, which offers an experimental opportunity to reveal the spatial configurations of the hedgehogs and antihedgehogs in real magnets such as MnSi$_{1-x}$Ge$_x$ and SrFeO$_3$. Our findings are also expected to provide insights into dynamics of hedgehog lattices in other types of magnets including insulating systems~\cite{Yang2016,Aoyama2021,Watanabe2021} and their couplings to other degrees of freedom such as lattices and polarizations~\cite{ZhangXX2017}. The nature and properties of the intrinsic excitation modes in the hedgehog lattices in magnets will open a new research field on the fundamental physics and even engineering of emergent magnetic monopoles in condensed matters.

This work was supported by JSPS KAKENHI (Grant No. 20H00337 and No. 23H04522), JST CREST (Grant No. JPMJCR20T1), and Waseda University Grant for Special Research Projects (2023C-140). R.E. was supported by a Grant-in-Aid for JSPS Fellows (Grant No. 23KJ2047). A part of the numerical simulations was carried out at the Supercomputer Center, Institute for Solid State Physics, University of Tokyo.

\pagebreak
\widetext
\begin{center}
\textbf{\large Supplemental Material for \\ “Theory of Collective Excitations in the Quadruple-Q Magnetic Hedgehog Lattices”}
\end{center}
\setcounter{equation}{0}
\setcounter{figure}{0}
\setcounter{table}{0}
\setcounter{page}{1}
\makeatletter
\renewcommand{\theequation}{S\arabic{equation}}
\renewcommand{\thefigure}{S\arabic{figure}}
\renewcommand{\thepage}{S\arabic{page}}
\tableofcontents
\section{Chemical potential dependence of the magnetic wavenumber}
In the present study, the chemical potential is fixed at a specific value of $\mu=-3.79$ to produce a commensurate magnetic wavenumber of $Q_{\rm abs}=\pi/4$ for the quadruple-$Q$ hedgehog lattices (4$Q$-HLs) for the convenience of numerical simulations with finite-size systems. When the chemical potential varies, the shape of Fermi surafce changes, which modulates the ordering vectors $\{\bm Q_\nu\}$. In this section, we discuss that the 4$Q$-HL states emerge not only at this specific value of $\mu$ but also within a rather wide range of $\mu$ by argument based on the bare magnetic susceptibility $\chi_{\bm q}^0$. By the second-order perturbation expansion with respect to the spin-charge coupling term (the second term) of the Kondo-lattice Hamiltonian $\mathcal{H}_{\rm KLM}$ in Eq.(1) in the main text, the free energy of the system is obtained as~\cite{Hayami2017},
\begin{align}
 F^{(2)} &= -\frac{J_{\rm K}^2}{4} \sum_{{\bm q}\in{\rm BZ}} \chi_{\bm q}^0 \left| {\bm S}_{\bm q} \right|^2, \label{F2}
\end{align}
with
\begin{align}
    \chi_{\bm q}^0 &= \frac{1}{N} \sum_{{\bm k}\in{\rm BZ}} \frac{f\left(\varepsilon_{\bm k}-\mu\right) - f\left(\varepsilon_{{\bm k}+{\bm q}}-\mu\right)}{\varepsilon_{{\bm k}+{\bm q}} - \varepsilon_{{\bm k}}}, \label{chiq}
\end{align}
where $f(\varepsilon)$ and $\varepsilon_{\bm k}$ denote the Fermi distribution function and the momentum representation of kinetic energy for the first term of $\mathcal{H}_{\rm KLM}$, respectively. The maxima of $\chi_{\bm q}^0$ give the propagation vectors of magnetic helices $\{{\bm Q}_\nu\}$. Here, $\nu$ is an index of multiple equivalent peaks of $\chi_{\bm q}^0$, if any, which are connected by symmetry operations of the cubic lattice.\par
\begin{figure}[tbh]
    \centering
    \includegraphics[scale=1.0]{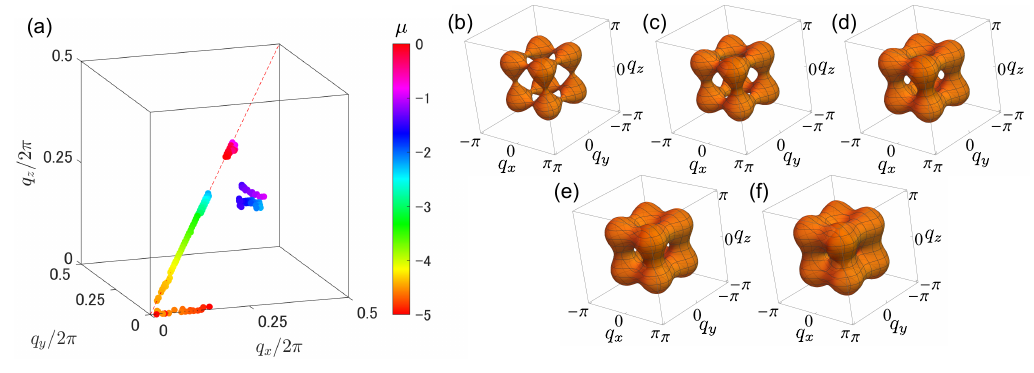}
    \caption{(a) Chemical potential dependence of the wavenumber ${\bm q}=(q_x,q_y,q_z)$ at which the bare magnetic susceptibility $\chi_{\bm q}^0$ has maximal peaks in the crystallographic Brillouin zone. The model parameters are set as $t_1=1$, $t_4=-1$, and the system size is $N=192^3$. The red line represents a high-symmetric momentum path which satisfies $q_x=q_y=q_z$. We present only one maximum for a given chemical potential $\mu$ which satisfies $q_x\geq q_y\geq q_z\geq 0$ for clear visibility, even when there are multiple maxima of $\chi_{\bm q}^0$. (b)-(f) Three-dimensional views of Fermi surfaces when (b) $\mu=-4.4$, (c) $\mu=-4.0$, (d) $\mu=-3.5$, (e) $\mu=-3.0$, and (f) $\mu=-2.4$.}
    \label{FigS01}
\end{figure}

We know from experience that the combination of the nearest-neighbor positive electron hopping and the farther-neighbor negative hopping often induces multiple equivalent maxima of $\chi_{\bm q}^0$ on the high-symmetric lines in the Brillouin zone~\cite{Hayami2017}. Here we set $t_1=1$ and $t_4=-1$, which is also presented in the main text. Fig.~\ref{FigS01}(a) shows chemical potential dependence of the ordering wavenumber at which $\chi_{\bm q}^0$ has maximal peaks. We find the wavenumber appears on a line of $q_x=q_y=q_z$ when $\mu$ is within a certain range of $-4.4\lesssim\mu\lesssim-2.4$. Note that the existence of the maximal peaks of $\chi_{\bm q}^0$ at such high-symmetric momenta is necessary and sufficient conditions for realization of the 4$Q$-geometry. This observation indicates that the 4$Q$-geometry is realized in the finite range of chemical potential. For further information, the Fermi-surface geometries for various values of chemical potential are presented in Fig.~\ref{FigS01}(b)-(f).
\section{Perturbational analysis for the effective fourth-order spin-spin interactions}
\begin{figure}[tbh]
    \centering
    \includegraphics[scale=1.0]{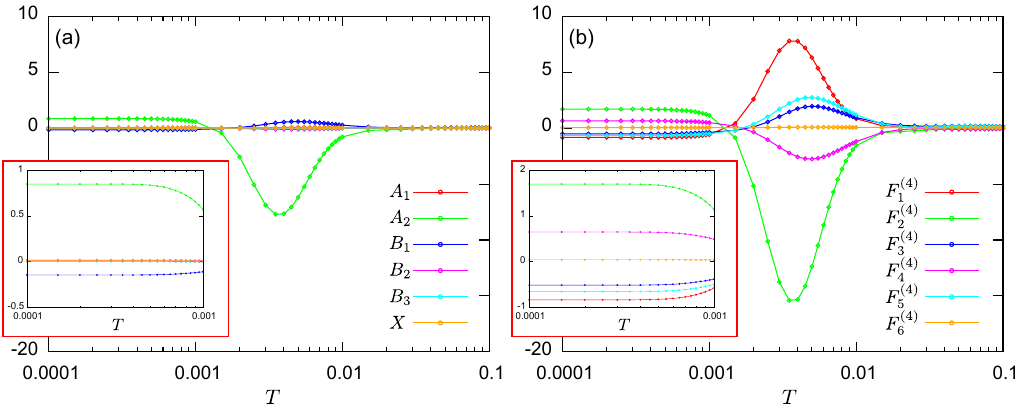}
    \caption{Numerically evaluated temperature-dependence of (a) the coefficients in Eqs.(\ref{eq:coeffA1})-(\ref{eq:coeffX}) and (b) the partial free energies from various scattering processes of the four-body spin-spin interactions in Eqs.(\ref{f41})-(\ref{f46}). Note that the magnitudes of partial free energies in (b) are normalized by $J_{\rm K}^4/(16N)$. Insets are magnified plots for lower-temperature region of (a) and (b), respectively. For the numerical calculations, we use  a system of $N=48^3$ sites and take $2^{20}$ points for the Matsubara frequencies.}
    \label{FigS02}
\end{figure}

The 4$Q$-geometry realized by the two-body spin-spin interactions discussed in the previous section is necessary but not sufficient to stabilize the 4$Q$-HL states. In order to stabilize the 4$Q$ magnetic structures, the four-body spin-spin interactions are indispensable, whose effects have been discussed in previous studies~\cite{Akagi2012,Hayami2014,Hayami2017}.
In this section, following a procedure presented in Ref.~\cite{Hayami2017}, we evaluate contributions of the effective fourth-order spin-spin interactions to the free energy by using the perturbational analysis of $\mathcal{H}_{\rm KLM}$. By sequential perturbation expansions with respect to the Kondo coupling in the weak-coupling regime of $J_{\rm K}\ll\max_{{\bm k}\nu}\left[\varepsilon_{{\bm k}\nu}-\mu\right]$~\cite{Hayami2017}, the effective fourth-order spin-spin interactions are obtained as,
\begin{align}
    F^{(4)} &= F^{(4)}_1 + F^{(4)}_2 + F^{(4)}_3 + F^{(4)}_4 + F^{(4)}_5 + F^{(4)}_6, \\
    F_1^{(4)} &= \frac{J_{\rm K}^4}{16N}\sum_\nu (2A_1-A_2) \left( {\bm S}_{{\bm Q}_\nu}\cdot{\bm S}_{{\bm Q}_\nu} \right) \left( {\bm S}_{-{\bm Q}_\nu}\cdot{\bm S}_{-{\bm Q}_\nu} \right), \label{f41} \\
    F_2^{(4)} &= \frac{J_{\rm K}^4}{16N} \sum_{\nu} (2A_2) \left( {\bm S}_{{\bm Q}_\nu}\cdot{\bm S}_{-{\bm Q}_\nu} \right)^2, \label{f42} \\
    F_3^{(4)} &= \frac{J_{\rm K}^4}{16N} \sum_{\nu,\nu'} 4(B_1+B_2-B_3) \left( {\bm S}_{{\bm Q}_\nu}\cdot{\bm S}_{-{\bm Q}_\nu} \right) \left( {\bm S}_{{\bm Q}_{\nu'}}\cdot{\bm S}_{-{\bm Q}_{\nu'}} \right), \label{f43} \\
    F_4^{(4)} &= \frac{J_{\rm K}^4}{16N} \sum_{\nu,\nu'} 4(-B_1+B_2+B_3) \left( {\bm S}_{{\bm Q}_\nu}\cdot{\bm S}_{{\bm Q}_{\nu'}} \right) \left( {\bm S}_{-{\bm Q}_\nu}\cdot{\bm S}_{-{\bm Q}_{\nu'}} \right), \label{f44} \\
    F_5^{(4)} &= \frac{J_{\rm K}^4}{16N} \sum_{\nu,\nu'} 4(B_1-B_2+B_3) \left( {\bm S}_{{\bm Q}_\nu}\cdot{\bm S}_{-{\bm Q}_{\nu'}} \right) \left( {\bm S}_{-{\bm Q}_{\nu'}}\cdot{\bm S}_{{\bm Q}_\nu} \right), \label{f45} \\
    F_6^{(4)} &= \frac{J_{\rm K}^4}{16N} 4X \left\{ \left( {\bm S}_{{\bm Q}_1}\cdot{\bm S}_{{\bm Q}_2} \right) \left( {\bm S}_{{\bm Q}_3}\cdot{\bm S}_{{\bm Q}_4} \right) \right. \notag \\
    &\quad\quad\quad \left. +\left( {\bm S}_{{\bm Q}_1}\cdot{\bm S}_{{\bm Q}_3} \right) \left( {\bm S}_{{\bm Q}_2}\cdot{\bm S}_{{\bm Q}_4} \right) + \left( {\bm S}_{{\bm Q}_1}\cdot{\bm S}_{{\bm Q}_4} \right) \left( {\bm S}_{{\bm Q}_2}\cdot{\bm S}_{{\bm Q}_3}\right) + {\rm c.c.} \right\}, \label{f46}
\end{align}
where the coefficients are given by products of noninteracting Green's functions $G_{\bm k}=\left(i\omega_p-\varepsilon_{\bm k}+\mu\right)^{-1}$ as,
\begin{align}
    A_1 &= \frac{T}{N} \sum_{{\bm k},\omega_p} (G_{\bm k})^2 G_{{\bm k}-{\bm Q}_\nu} G_{{\bm k}+{\bm Q}_\nu},
    \label{eq:coeffA1} \\
    A_2 &= \frac{T}{N} \sum_{{\bm k},\omega_p} (G_{\bm k})^2 (G_{{\bm k}+{\bm Q}_\nu})^2, \\
    B_1 &= \frac{T}{N} \sum_{{\bm k},\omega_p} (G_{\bm k})^2 G_{{\bm k}+{\bm Q}_\nu} G_{{\bm k}+{\bm Q}_{\nu'}}, \\
    B_2 &= \frac{T}{N} \sum_{{\bm k},\omega_p} (G_{\bm k})^2 G_{{\bm k}+{\bm Q}_\nu} G_{{\bm k}-{\bm Q}_{\nu'}}, \\
    B_3 &= \frac{T}{N} \sum_{{\bm k},\omega_p} G_{\bm k} G_{{\bm k}+{\bm Q}_\nu} G_{{\bm k}+{\bm Q}_{\nu'}} G_{{\bm k}+{\bm Q}_\nu+{\bm Q}_{\nu'}}, \\
    X   &= \frac{T}{N} \sum_{{\bm k},\omega_p} G_{\bm k} G_{{\bm k}+{\bm Q}_1} G_{{\bm k}+{\bm Q}_1+{\bm Q}_2} G_{{\bm k}+{\bm Q}_1+{\bm Q}_2+{\bm Q}_3}. \label{eq:coeffX}
\end{align}
Here $\omega_p$ denotes the fermionic Matsubara frequency. Fig.~\ref{FigS02} shows temperature-dependence of each coefficient given by Eqs.(\ref{eq:coeffA1})-(\ref{eq:coeffX}) and partial free energy from each scattering process given by Eqs.(\ref{f41})-(\ref{f46}) for the parameter set  used for the present study, i.e., $(t_1,t_4,\mu)=(1,-1,-3.79)$. As seen in Fig.~\ref{FigS02}(a), the coefficient $A_2$ has the largest absolute value than any other coefficients at low temperatures. Hence the contribution of $F_2^{(4)}$ to the free energy is positive and most significant at low temperatures among all the four-body spin-spin interaction terms. Combining this result about the four-body spin-spin interactions with the knowledge about the two-body spin-spin interactions discussed in the previous section, we find that an effective model of the Kondo-lattice model examined in the present study is the bilinear-biquadratic model, whose Hamiltonian is given by,
\begin{equation}
    \mathcal{H}_{\rm eff} = 2\sum_\nu \left[ -J|{\bm S}_{{\bm Q}_\nu}|^2 + \frac{K}{N}|{\bm S}_{{\bm Q}_\nu}|^4 \right].
\end{equation}
The ground-state spin configuration for this effective model was studied in the previous work \cite{Okumura2020}. In this Hamiltonian, only the contributions of $F^{(4)}_2$ and $F^{(2)}$ at the ordering vectors  $\{\bm Q_\nu \}$ ($\nu=1,2,3,4$) are taken into account in a phenomenological manner.  Even with this simplification, this effective spin model reproduces the zero-field ground state of the original Kondo-lattice model, i.e., the 4$Q$-HL state. 
\section{Phase diagram}
\begin{figure}[tbh]
\centering
\includegraphics[scale=1.0]{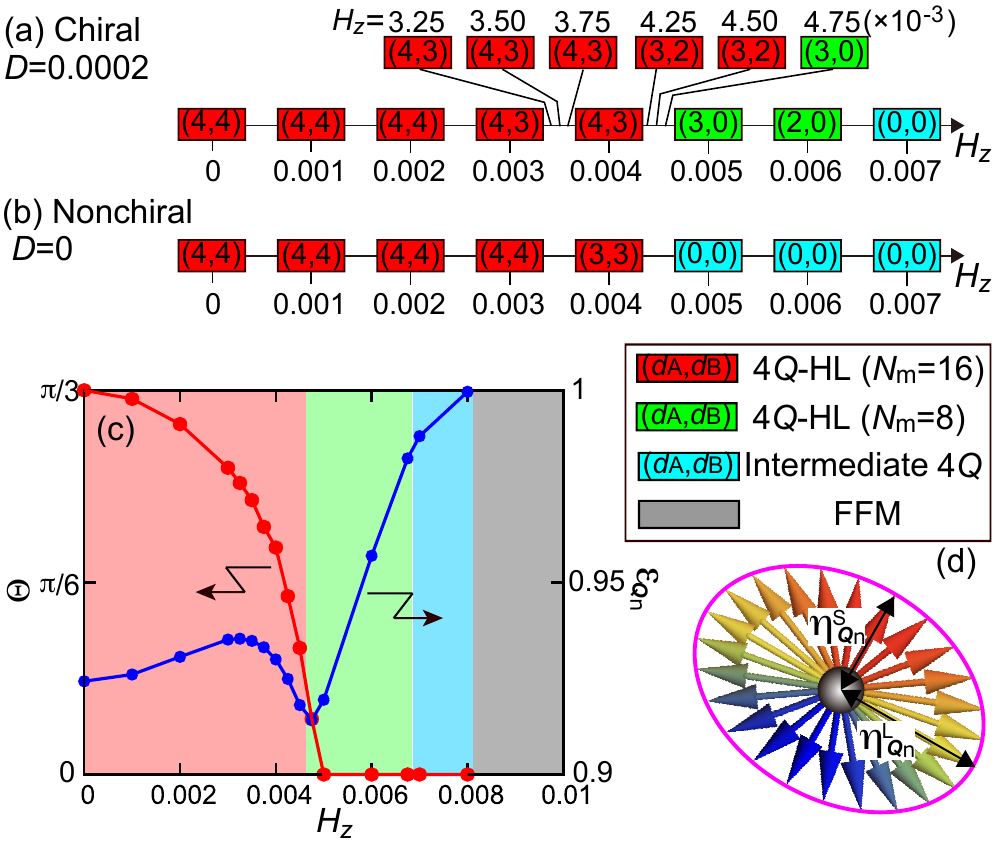}
\caption{(a),~(b) Detailed phase diagram of $\mathcal{H}$ in the main text as a function of the external magnetic field $H_z$ for (a) the chiral case with $D=0.0002$ and (b) the nonchiral case with $D=0$. The pair of numbers $(d_{\rm A},d_{\rm B})$ in each box represents the lengths of Dirac strings A and B, and $d_{\rm A}=0$ ($d_{\rm B}=0$) means the Dirac strings A (B) do not exist or the length of Dirac strings A (B) is shorter than the lattice constant. (c) Field dependence of the phase degree of freedom $\Theta$ and the ellipticity of four helices $\varepsilon_{{\bm Q}_\nu}$ constituting the 4$Q$-HL spin structure. Since the ellipticities of the four helices are equivalent for a given magnetic field, we simply denote $\varepsilon_{{\bm Q}_n}(=\varepsilon_{{\bm Q}_1}=\varepsilon_{{\bm Q}_2}=\varepsilon_{{\bm Q}_3}=\varepsilon_{{\bm Q}_4})$. The light pink, light green, light blue, and gray backgrounds show the 4$Q$-HL phase ($N_{\rm m}$=16), the 4$Q$-HL phase ($N_{\rm m}$=8), the intermediate 4$Q$ phase, and the forced ferromagnetic phase, respectively. (d) Helix of the localized spins seen along the propagation vector. The ellipticity of the helix is defined by the ratio between lengths of the short and long axes as $\varepsilon_{{\bm Q}_n}=\eta_{{\bm Q}_n}^{\rm S}/\eta_{{\bm Q}_n}^{\rm L}$.}
\label{FigS03}
\end{figure}
Figures~\ref{FigS03}(a) and \ref{FigS03}(b) show phase diagrams of $\mathcal{H}$ in the main text as a function of the external magnetic field $H_z$ for the chiral case with $D=0.0002$ and the nonchiral case with $D=0$, respectively. In the nonchiral case with $D=0$, the zero-field ground state is the 4$Q$-HL with $N_{\rm m}=16$, which is composed of four vortex-type Dirac strings A and four antivortex-type Dirac strings B in its magnetic unit cell. This 4$Q$-HL spin structure can be interpreted as a superposition of two helices with lefthanded chirality and two helices with righthanded chirality~\cite{Okumura2022}. In other words, pairs of helices with opposite chiralities constitute the nonchiral 4$Q$-HL. This pairwise chiralities of helices result in the cancellation of net scalar chirality for the nonchiral 4$Q$-HL state even in the presence of a magnetic field. As the magnetic field increases, the Dirac strings are shortened gradually and eventually the hedgehog and antihedgehog belonging to the string A share the same unit cube and, consequently, a transition from the nonchiral 4$Q$-HL phase with $N_{\rm m}=16$ to an intermediate 4$Q$ phase occurs. It is worth mentioning that the Dirac strings A and B always have equal lengths ($d_{\rm A}=d_{\rm B}$) upon the variation of external magnetic field $H_z$.\par
In the chiral case with $D=0.0002$ in Fig.~\ref{FigS03}(a), the zero-field ground state is again the 4$Q$-HL with $N_{\rm m}=16$ composed of four strings A and four strings B in the magnetic unit cell. However, the spin structure of this 4$Q$-HL is distinct from that of the nonchiral case. The chiral 4$Q$-HL state can be regarded as a superposition of four righthanded helices and thus has a finite scalar chirality in the presence of magnetic field~\cite{Okumura2020}. As the magnetic field $H_z$ increases, four Dirac strings A and four Dirac strings B in the magnetic unit cell are shortened gradually. Here, the shortening rate of the strings B is faster than that of the strings A. Eventually the first topological transition occurs from the 4$Q$-HL phase with $N_{\rm m}=16$ to a new 4$Q$-HL phase with $N_{\rm m}=8$ when the strings B disappear due to the hedgehog-antihedgehog pair annihilations at the strings B. Further increase of the magnetic field leads to further shortening of the strings A. Finally, the second transition occurs from the 4$Q$-HL phase with $N_{\rm m}=8$ to the intermediate 4$Q$ phase when the hedgehog and antihedgehog belonging to the string A share the same unit cube, as seen in Fig.~\ref{FigS07}(d). As further increase in the magnetic field, the intermediate 4$Q$ phase turns into the forced ferromagnetic phase [See also Fig.~\ref{FigS07}(e)] with an abrupt magnetization jump presented in Fig.~\ref{FigS08}. \par
The first topological transition is characterized by abrupt changes of internal degrees of freedom in the chiral 4$Q$-HL state. We focus on the phase degree of freedom $\Theta$ shown in Fig.~\ref{FigS03}(c) which is given by the sum of the translational degree of freedom of four helices $\theta_\nu$ $(\nu=1,2,3,4)$ as described in the main text. In the absence of magnetic field, $\Theta$ is fixed at $\pi/3$. As the magnetic field increases, $\Theta$ decreases gradually and exhibits abrupt decrease when $H_z$ reaches aroud the threshold field. Eventually, $\Theta$ goes to zero when $H_z$ slightly exceeds the threshold field. More precisely, the first topological transition occurs shortly before $\Theta$ goes to zero, which can be understood from the results of previous studies treating a spatially continuum limit~\cite{Shimizu2022}. We can also see that the ellipticities of the four helices $\varepsilon_{{\bm Q}_n}=\eta_{{\bm Q}_n}^{\rm S}/\eta_{{\bm Q}_n}^{\rm L}$ [Fig.~\ref{FigS03}(d)] shows an anomalous behavior near the transition point. The ellipticity usually increases as the magnetic field increases, but it decreases near the transition point.
\section{Mode localizations in the hedgehog lattice phases}
\begin{figure}[tbh]
\centering
\includegraphics[scale=1.25]{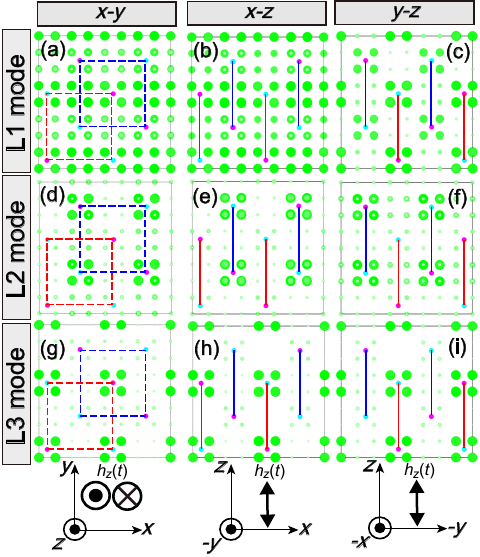}
\caption{Similar plot as Fig. 3 in the main text for the hedgehog lattice phase with $N_{\rm m}=16$ at $H_z=0$.}
\label{FigS04}
\end{figure}
\begin{figure}[tbh]
\centering
\includegraphics[scale=1.25]{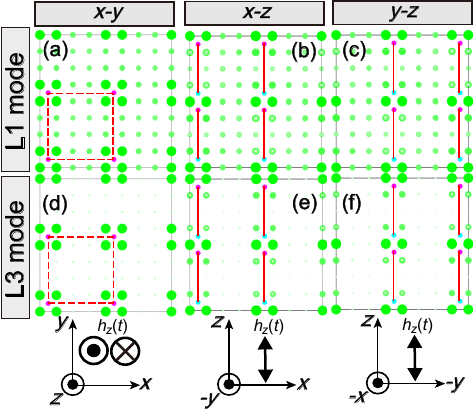}
\caption{Similar plot as Fig. 3 in the main text for the hedgehog lattice phase with $N_{\rm m}=8$ at $H_z=0.005$.}
\label{FigS05}
\end{figure}
In this section, we present similar data as Fig. 3 in the main text for different values of external magnetic fields. Figs.~\ref{FigS04} and \ref{FigS05} show the oscillation amplitudes of the modes in the $N_{\rm m}=16$ at $H_z=0$ and those in the $N_{\rm m}=8$ phase at $H_z=0.005$, respectively. In these figures, we can see almost similar spatial distributions of oscillation amplitudes as Fig. 3 in the main text for each mode, where the L1 mode is distributed throughout the magnetic unit cell and the L2 (L3) mode is localized around the strings B (A). Note that we do not have the data of the L2 mode in Fig.~\ref{FigS05} because the L2 mode is absent in the $N_{\rm m}=8$ phase. It is also worth mentioning that the $C_{4z}$-breaking nature seen in Figs.~\ref{FigS04}(b) and \ref{FigS04}(c) and the $C_{4z}$-preserving nature seen in Figs.~\ref{FigS05}(b) and \ref{FigS05}(c) are closely related to the symmetry nature of spin textures discussed in Sec.~VI.
\section{String dynamics in the L1 mode}
\begin{figure}[tbh]
\centering
\includegraphics[scale=0.95]{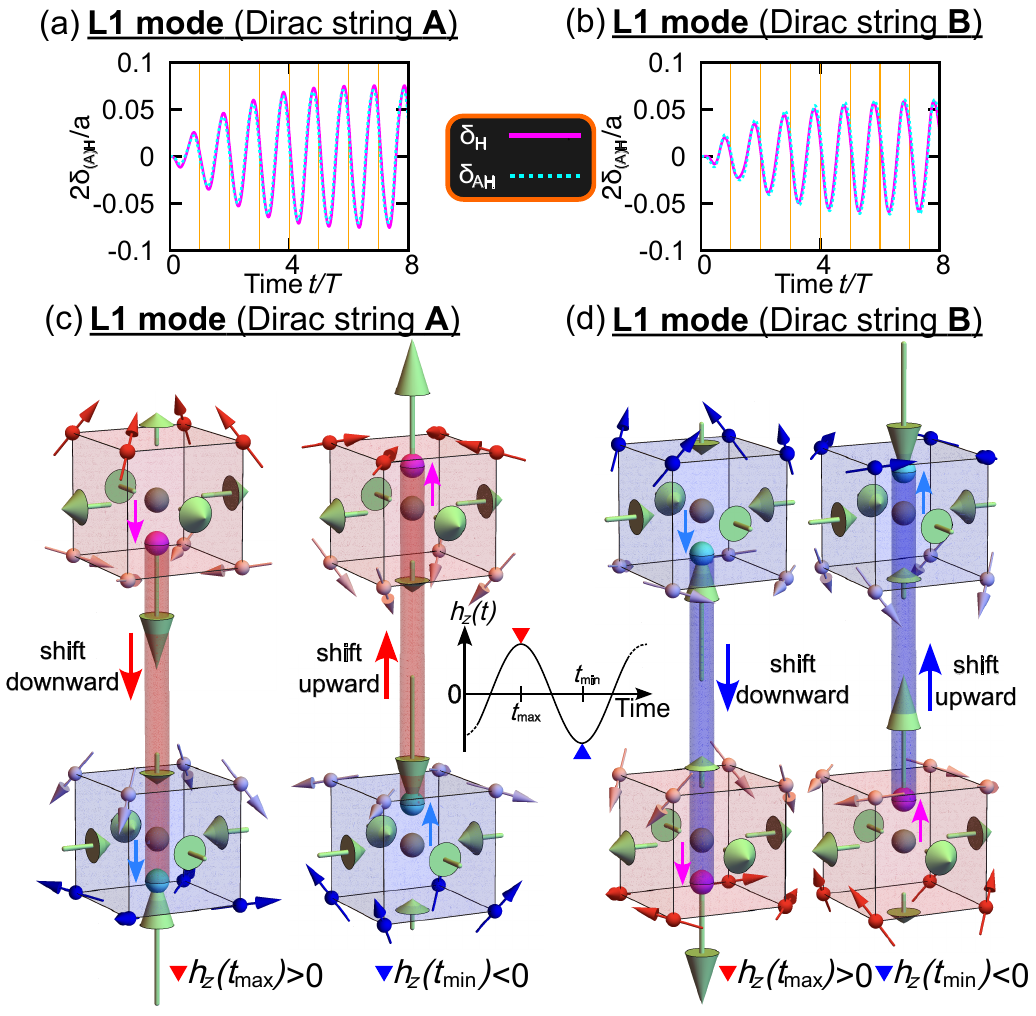}
\caption{(a), ~(b) Time profiles of the displacements $\delta_{\rm (A)H}$ of the Bloch points in the L1 mode for (a) String A and (b) String B. (c),~(d) Schematics of the oscillatory translational motion of the Dirac strings in the L1 mode for (c) String A and (d) String B. Note that $T=2\pi/\omega_{\rm ac}$ is the time periodicity of the ac magnetic field $h_z(t)$.}
\label{FigS06}
\end{figure}
While the L1 mode is not localized around the strings as explained in the main text, we find finite oscillation amplitudes around both the Strings A and B when we activate the L1 mode by applying an ac magnetic field with the corresponding frequency. So, regardless of how meaningful the following analysis is, we can extract the dynamics of both Strings A and B in the L1 mode by using the same method for the L2 and L3 modes. Figures.~\ref{FigS06}(a) and \ref{FigS06}(b) show calculated time profiles of displacements $\delta_{\rm (A)H}$ for the Strings A and B in the L1 mode. For both strings, $\delta_{\rm H}$ and $\delta_{\rm AH}$ show in-phase time-periodic oscillations, which indicates translational motion of the Dirac strings. Schematics of this translational motion is shown in Fig.~\ref{FigS06}(c) and \ref{FigS06}(d). The oscillation phases of both Strings A and B in the L1 mode are opposite to those in the L2 and L3 modes. In contrast to the dynamics of Strings B (A) in the L2 (L3) mode which shifts upward (downward) when the sign of the ac magnetic field $h_z(t)$ is positive (negative) as argued in the main text, both the Strings A and B in the L1 mode shift downward (upward) when the sign of $h_z(t)$ is positive (negative).
\section{Intermediate 4$Q$ phase}
\begin{figure}[tbh]
\centering
\includegraphics[scale=0.5]{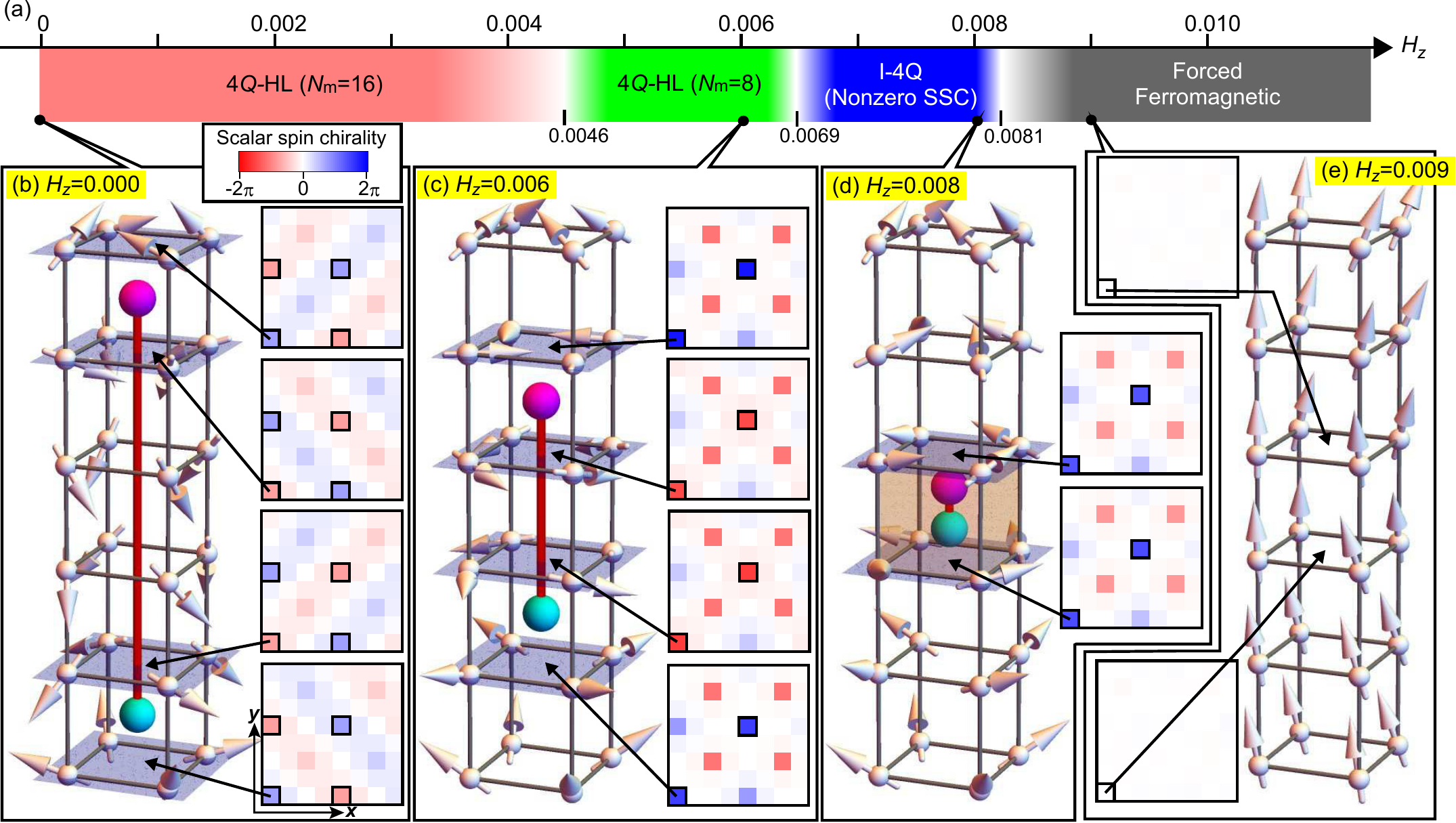}
\caption{Calculated spatial configurations of the spins around the Dirac string A and spatial maps of the local scalar spin chirality on the $xy$ planes (8$\times$8-site portion of the 16$\times$16-site planes) at selected $z$ coordinates for respective phases in the phase diagram, i.e., (a) the $4Q$-HL phase with $N_{\rm m}$=16, (b) the $4Q$-HL phase with $N_{\rm m}$=8, (c) the intermediate $4Q$ (I-4$Q$) phase, and (d) the forced ferromagnetic phase.}
\label{FigS07}
\end{figure}
\begin{figure}[tbh]
\centering
\includegraphics[scale=0.8]{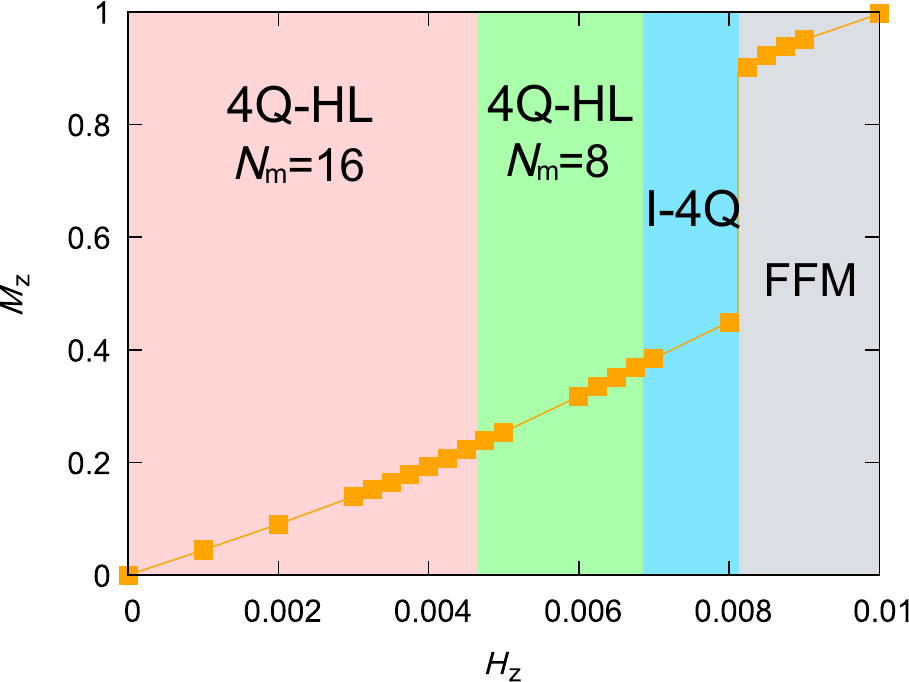}
\caption{Calculated magnetization profile as a function of $H_z$.}
\label{FigS08}
\end{figure}
In the intermediate 4$Q$ phase (I-4$Q$ phase) in the phase diagram (Fig.~2 in the main text), the hedgehog and antihedgehog connected by a Dirac string A collide to be merged, and their cores share the same cubic unit cell. This phase has nonzero scalar spin chirality around the merging point although it no longer takes a quantized value. Figure~\ref{FigS07} shows the calculated spatial spin configurations around the Dirac string A and the spatial maps of  local scalar spin chirality on the $xy$ planes at selected $z$ coordinates for respective phases. We find that the nonzero scalar spin chirality indeed remains in the I-4$Q$ phase around the hedgehog-antihedgehog merging point. In Fig.~2 in the main text, we find that the L3 mode associated with the oscillation of Dirac strings A survives in the I-4$Q$ phase. This L3 mode originates from residues of  the hedgehogs and antihedgehogs as  a remnant of magnetic topology.
The nonzero scalar spin chirality in the I-4$Q$ phase survives and takes a rather large value even on the verge of the phase boundary to the forced ferromagnetic phase with zero scalar spin chirality. This is because the phase transition between the I-4$Q$ phase and the forced ferromagnetic (FFM) phase is of strong first-order. Indeed, in Fig.~\ref{FigS07}, we can see a nearly 90-degree flop of spins from in-plane to out-of-plane directions and abrupt vanishing of scalar spin chirality when the system enters the FFM. This first-order nature of the transition can also be seen in the calculated magnetization profile as a function of $H_z$ in Fig.\ref{FigS08}, which shows an abrupt jump at the transition point.
\section{Remarks on the inherent $C_4$-symmetry breaking}
In Figs. 3(b) and 3(c) in the main text, we find the breaking of $C_{4z}$ symmetry in the spatial distribution of the spin-oscillation magnitudes in the $N_{\rm m}=16$ phase. The absence of $C_{4z}$ symmetry is attributable to the breaking of $\tilde{C}_{4z}$ symmetry in the spin configuration in equilibrium despite the Hamiltonian always preserves the $C_{4z}$ symmetry. In this section, we use the term ``$\tilde{C}_{4}$ symmetry" with tilde in a broader sense than we generally mean by the term ``$C_4$ symmetry". Specifically, while the  term ``$C_4$ symmetry" in magnetically ordered systems usually means that the spatial spin configuration is invariant under a fourfold rotation around a particular axis, ``$\tilde{C}_4$ symmetry" in this section means merely the invariance of the absolute values of (static) spin structure factors $S_{\bm q}=\left(1/\sqrt{N}\right)\sum_i {\bm S}_i e^{i{\bm q}\cdot{\bm r_i}}$ under the fourfold rotation. Namely the term does not necessarily mean the invariance of the spin configuration. The magnetic ordering with broken $\tilde{C}_4$ symmetry appears even in the absence of magnetic field [Figs.~\ref{FigS04}(b) and \ref{FigS04}(c)], where only one of the three $\tilde{C}_4$ symmetries ($\tilde{C}_{4x}$, $\tilde{C}_{4y}$, and $\tilde{C}_{4z}$) is preserved and the other two are broken.  In the case of Figs.~\ref{FigS04}(a)-(c), for example, the $\tilde{C}_{4x}$ symmetry is preserved whereas the $\tilde{C}_{4y}$ and $\tilde{C}_{4z}$ symmetries are broken. Here, two axes for the broken $\tilde{C}_4$ symmetries are equivalent. 

In the presence of  external magnetic field, the situation is more complicated. This is because the spin configuration obtained by adiabatic application of magnetic field to the zero-field magnetic texture with one $\tilde{C}_{4}$-symmey and two broken $\tilde{C}_{4}$-symmetries differs depending on whether the magnetic field is applied along the $\tilde{C}_{4}$-axis for  the prserved symmetry or not. In the hedgehog lattice phase with $N_{\rm m}=16$ in the presence of magnetic field, the spin configuration obtained by applying a magnetic field along the direction parallel to one of the two axes for the broken symmetries at zero field is the ground-state spin configuration because it is lower in energy than that obtained by applying a magnetic field with the same magnitude along the unique direction parallel to the axis for the preserved  symmetry. Consequently, the $\tilde{C}_{4z}$ symmetry is always broken in the hedgehog-lattice phase with $N_{\rm m}=16$ in the presence of finite magnetic field along the $z$-axis because we consider only the ground-state spin configurations but not metastable ones in this work. To obtain the ground-state spin configurations, we performed an unbiased numerical analysis based on the kernel polynomial method in the present work. 

In addition, the $x$-, $y$-, and $z$-axes in the absence of magnetic field are defined to be consistent with the situation in a magnetic field along the $z$-axis . Thereby, the $\tilde{C}_{4}$-symmetry axis at zero magnetic field should not be the $z$-axis but either $x$-axis or $y$-axis. Thus, the $\tilde{C}_{4z}$ symmetry is always broken in the $N_{\rm m}=16$ phase treated in the present work irrespective of the presence or absence of the magnetic field along the $z$-axis. We also note that the $\tilde{C}_{4z}$ symmetry which is broken in the $N_{\rm m}=16$ phase is abruptly recovered when the system enters the $N_{\rm m}=8$ phase [Figs.~\ref{FigS05}(b) and \ref{FigS05}(c)]. The inherent breaking of $\tilde{C}_{4z}$ symmetry in the hedgehog-lattice spin configurations might be related with the phase degree of freedom $\Theta$. Indeed, $\Theta$ is finite in the $N_{\rm m}=16$ phase with broken $\tilde{C}_{4z}$ symmetry, while it is zero in the $N_{\rm m}=8$ phase with $\tilde{C}_{4z}$ symmetry [Fig.~\ref{FigS03}(c)]. To clarify the relation between $\Theta$ and the symmetry breaking of the hedgehog lattice spin configurations is left for future study.

\end{document}